\documentclass[aps,
amssymb,
amsmath,
superscriptaddress,
reprint]{revtex4-2}
\usepackage{graphicx}
\usepackage{dcolumn}
\usepackage{bm}

\usepackage[utf8]{inputenc}
\usepackage{times}
\usepackage{siunitx}
\usepackage{booktabs}
\usepackage{graphicx}
\usepackage{lineno}
\usepackage{comment}

\def\SI#1{\textcolor{black}{#1}}

\begin{document}


\title{Light-induced torque in ferromagnetic metals via orbital angular momentum generated by photon-helicity}
\author{Koki Nukui}
\affiliation{Department of Applied Physics, Graduate School of Engineering, Tohoku University, Sendai 980-8579, Japan}
\affiliation{WPI Advanced Institute for Materials Research (AIMR), Tohoku University, 2-1-1, Katahira, Sendai 980-8577, Japan}

\author{Satoshi Iihama}
\email{iihama.satoshi.y0@f.mail.nagoya-u.ac.jp, (present address: Department of Materials Physics, Nagoya University, Nagoya 464-8603, Japan)}
\affiliation{Frontier Research Institute for Interdisciplinary Sciences (FRIS), Tohoku University, Sendai 980-8578, Japan}
\affiliation{WPI Advanced Institute for Materials Research (AIMR), Tohoku University, 2-1-1, Katahira, Sendai 980-8577, Japan}

\author{Kazuaki Ishibashi}
\affiliation{Department of Applied Physics, Graduate School of Engineering, Tohoku University, Sendai 980-8579, Japan}
\affiliation{WPI Advanced Institute for Materials Research (AIMR), Tohoku University, 2-1-1, Katahira, Sendai 980-8577, Japan}

\author{Shogo Yamashita}
\affiliation{Department of Applied Physics, Graduate School of Engineering, Tohoku University, Sendai 980-8579, Japan}

\author{Akimasa Sakuma}
\affiliation{Department of Applied Physics, Graduate School of Engineering, Tohoku University, Sendai 980-8579, Japan}

\author{Philippe Scheid}
\affiliation{Université de Lorraine, CNRS, Institut Jean Lamour, F-54000 Nancy, France}

\author{Grégory Malinowski}
\affiliation{Université de Lorraine, CNRS, Institut Jean Lamour, F-54000 Nancy, France}

\author{Michel Hehn}
\affiliation{Université de Lorraine, CNRS, Institut Jean Lamour, F-54000 Nancy, France}
\affiliation{Center for Science and Innovation in Spintronics (CSIS), Tohoku University, Sendai 980-8577, Japan}

\author{Stéphane Mangin}
\affiliation{Université de Lorraine, CNRS, Institut Jean Lamour, F-54000 Nancy, France}
\affiliation{Center for Science and Innovation in Spintronics (CSIS), Tohoku University, Sendai 980-8577, Japan}

\author{Shigemi Mizukami}
\email{shigemi.mizukami.a7@tohoku.ac.jp}
\affiliation{WPI Advanced Institute for Materials Research (AIMR), Tohoku University, 2-1-1, Katahira, Sendai 980-8577, Japan}
\affiliation{Center for Science and Innovation in Spintronics (CSIS), Tohoku University, Sendai 980-8577, Japan}

\date{\today}

\begin{abstract}

    We investigated photon-helicity-induced magnetization precession in Co$_{1-x}$Pt$_{x}$ alloy thin films. 
    In addition to field-like torque, attributable to magnetic field generation owing to {\it the inverse Faraday effect}, we observed non-trivial and large damping-like torque which has never been discussed for single ferromagnetic layer.
    The composition dependence of those two torques is effectively elucidated by a model that considers mutual coupling via spin-orbit interaction between magnetization and the electronic orbital angular momentum generated by photon-helicity. 
    This work significantly enhances our understanding of the physics relevant to the interplay of photon-helicity and magnetization in magnetic metals.
    
\end{abstract}

\maketitle

Photon carries both spin and orbital angular momenta, in circularly-polarized and vortex electromagnetic wave, respectively, and utilization of those angular momentum of photon is one issue in modern physics\cite{Bliokh2015b, Bliokh2015}. 
Considerable attention has been devoted in condensed matter physics to the interaction between photon's angular momentum and matter, fostering technological advances such as vortex-nano-processing\cite{Omatsu2010, Toyoda2012} and optical information communications\cite{Willner2021, Ji2020, Ishihara2023}. 

Intriguing physical phenomena relevant to the angular momentum of photon is also discussed as one of the central subjects in opto-magnetism. 
{\it The inverse Faraday effect} is a physical phenomenon in which magnetization ${\bf M}_{\rm p}$ \SI{or magnetic field ${\bf H}_{\rm p}$} emerges during the dwell time when circularly-polarized light passes through a dielectric medium, and it can be expressed as,
\SI{
\begin{align}
    {\bf M}_{\rm p}, {\bf H}_{\rm p} \propto {\bf E}(\omega _{\rm p})\times {\bf E}^{\star }(\omega _{\rm p}), \label{eq:Mp}
\end{align}}
where ${\bf E}$ is electric field vector for light with an angular frequency of $\omega _{\rm p}$\cite{Ziel1965, Pershan1966}. 
\SI{${\bf H}_{\rm p}$ can be generated instantaneously when light is irradiated on magnetic dielectric materials owing to magneto-optical coupling such as the magneto-optical Faraday effect\cite{Kirilyuk2010}.} 
Consequently, GHz-to-THz dynamics of magnetic order are initiated when a circularly-polarized femtosecond laser pulse passes through magnetic materials via the inverse Faraday effect, enabling us to explore various spin dynamics along with various probe, which is an ultrafast and versatile modern technique established to date\cite{Kimel2005, Satoh2010, Satoh2012, Choi2017, Savochkin2017, Tzschaschel2020}.

The inverse Faraday effect has also been discussed in magnetic metals and a highlight is circularly-polarized laser pulse-induced magnetization reversal, commonly referred to as All-Optical Helicity Dependent Switching (AO-HDS), partially attributed to the inverse Faraday effect\cite{Lambert2014, ElHadri2016, Medapalli2017, Takahashi2016, John2017, Kichin2019, Cheng2020}.
Considerable efforts have been devoted to understanding the physics of the inverse Faraday effect in metals; however, there is still a notable gap in our understanding regarding of the key physics: the nature of magnetization arising from the inverse Faraday effect in metallic ferromagnets. 
A recent study discussed the inverse Faraday effect in terms of the above-mentioned effective magnetic field to interpret the light-induced precessional magnetization dynamics observed in various ferromagnets and normal metal bilayers\cite{Choi2017, Choi2019, Choi2020, Iihama2021, Iihama2022, Wang2023}.
Many theories reported light-induced magnetization of metals\cite{Berritta2016, Hertel2006, Popova2012, Mondal2015, Freimuth2016, Li2017, Scheid2019, Mishra2023, Scheid2023}.
According to the recent theoretical analysis of the inverse Faraday effect magnetization ${\bf M}_{\rm p}$ in metals predominantly consists of {\it orbital angular momentum (OAM) and not spin angular momentum (SAM)}\cite{Berritta2016}.
On the other hand, in addition to the inverse Faraday effect, both {\it OAM} and {\it SAM} can be induced as an absorbed quantity\cite{Scheid2019, Scheid2021, Scheid2022, Scheid2023}.
Although the inverse Faraday effect is proportional to the laser intensity, {\it i.e.}, the effect vanishes after the laser passes away from the matter, {\it OAM} and {\it SAM} are induced by the absorption of photonic angular momentum that remains in matter even after the laser passes away from it. 
By using the absorption, {\it OAM} can be induced in any dissipative materials.
It is then intriguing to determine whether the light-induced angular momentum in metals is relevant to orbital magnetism in the context of recent research on the magnetization dynamics induced by an electric field (current) via {\it OAM}\cite{Go2020PRR1, Go2020PRR2, Ding2020, Lee2021CommunPhys, Lee2021NCommun, Choi2023, Hayashi2023}. 
Furthermore, their deep understanding will also push technologies relevant to opto-magnetism, such as the aforementioned AO-HDS.

\begin{figure*}[ht]
    \begin{center}
    \includegraphics[width=14cm,keepaspectratio,clip]{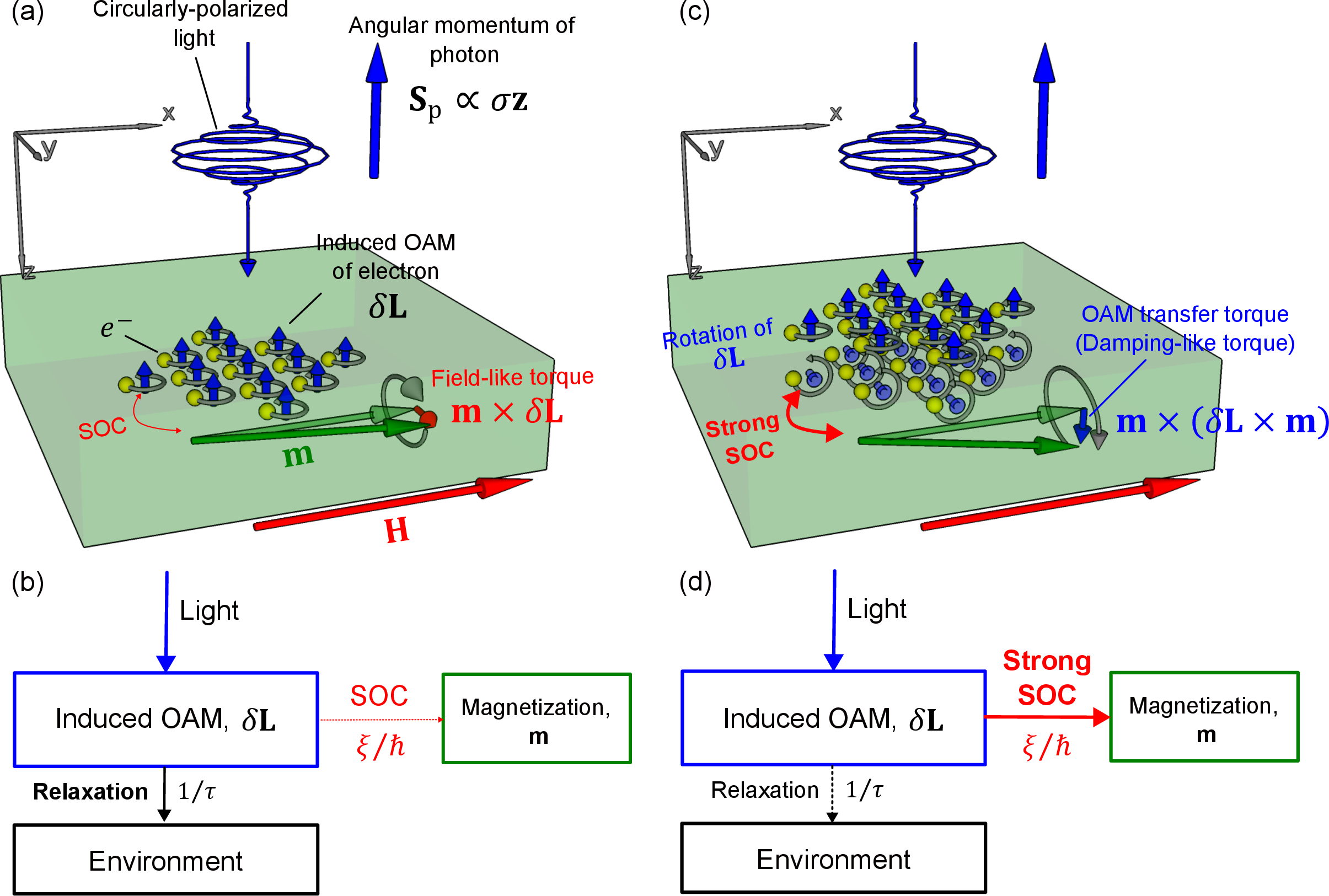}
    \end{center}
    \caption{ (a) and (c)  Schematic illustration of photon-helicity generated OAM of electron and magnetization dynamics excitation. 
    Circularly-polarized light has an angular momentum where its sign depends on helicity ($\sigma $). 
    The OAM ($\delta {\bf L}$) is generated parallel to ${\bf S}_{\rm p}$. 
    Magnetization dynamics are excited by SOC. 
    Schematic of angular momentum flow induced by photon-helicity when (b) relaxation to the environment is stronger than SOC and when (d) SOC is stronger than relaxation to environment. 
    When angular momentum flow through SOC dominates over relaxation to environment, $\delta {\bf L}$ rotates within laser pulse duration; then, magnetization dynamics can be excited by OAM transfer torque as shown in Fig. (c). }
    \label{f1}
\end{figure*}

In this Letter, we tend to elucidate the role of electronic OAM in the dynamics of magnetization induced by circularly-polarized laser pulses in ferromagnets. 
Our experimental data clearly demonstrates the existence of not only the effective magnetic field, traditionally interpreted as the inverse Faraday effect, but also a non-trivial and large effective magnetic field which is orthogonal to that induced by the inverse Faraday effect. 
In other words, these correspond to the field- and damping-like torques, respectively. 
In the former scenario circularly-polarized laser pulse acts as an effective magnetic field which induces field-like torque. 
On the other hand, the latter scenario where damping-like torque is generated cannot be explained by the traditional inverse Faraday effect alone. 
Both torques can be understood with a simple model provided, on an equal footing, considering the non-equilibrium OAM induced by the angular momentum of photon and its coupling with magnetization via spin-orbit interactions in ferromagnetic metals.
\SI{Here, it should also be noted that the effective magnetic field generation in the model is different with the traditional model [Eq. (1)] where magneto-optical coupling plays an important role.}   

As mentioned above, we focus on both kinds of torque (or effective magnetic field), and these are illustrated in Fig. 1. 
In Fig. 1 the macroscopic magnetization of a ferromagnetic metal film is in the film plane (${\bf m} \parallel {\bf x}$) and circularly-polarized laser pulse with ${\bf k}$ vector incidents along the film normal (${\bf k}\parallel {\bf z}$). 
In this geometry, magnetization precession is induced and the initial phase is determined by the direction of the total torque.
The \SI{absorption of photonic angular momentum} induces ${\bf M}_{\rm p}$ parallel to ${\bf z}$, which will then act on magnetization as an effective magnetic field, {\it i.e.}, the field-like torque, \SI{via a coupling between ${\bf m}$ and ${\bf M}_{\rm p}$}, ${\bf T} \propto {\bf m}\times {\bf M}_{\rm p}$ [Fig. \ref{f1}(a)], as reported\cite{Choi2017, Choi2019, Iihama2022}.
In the opposite case, the direction of the ${\bf M}_{\rm p}$ induced within a laser pulse duration changes owing to strong coupling between ${\bf M}_{\rm p}$ and ${\bf m}$, leading to the emergence of the damping-like torque, {\it i.e.}, ${\bf T}\propto {\bf m}\times ({\bf M}_{\rm p}\times {\bf m})$ [Fig. \ref{f1}(c)].
Hereafter, we assume that ${\bf M}_{\rm p}$ is dominantly caused by the electronic OAM $\delta {\bf L}$ induced by light, as shown in Fig. \ref{f1}.
The path of the angular momentum flow is schematically shown in Figs. 1(b) and 1(d). 
When relaxation to environment for induced OAM is stronger than the coupling with magnetization, light acts as the field-like torque [Figs. 1(a) and (b)]. 
However, coupling between induced OAM and magnetization is stronger than relaxation to environment, the damping-like torque as an OAM transfer torque is induced by light irradiation [Figs. 1(c) and 1(d)]. 
Consequently, the two cases are characterized by the coupling strength between ${\bf m}$ and $\delta {\bf L}$ and relaxation of $\delta {\bf L}$. 
In the realm of torque physics induced by the OAM generated by electric current or field, one can intuitively infer that the coupling strength between ${\bf m}$ and $\delta {\bf L}$ is governed by SOC, given that magnetization stems from SAM in transition metal ferromagnets. 
Hence, the field- and damping-like torques can be tuned by the heavy-metal addition into ferromagnetic metals.

\begin{figure}[ht]
    \begin{center}
    \includegraphics[width=8.5cm,keepaspectratio,clip]{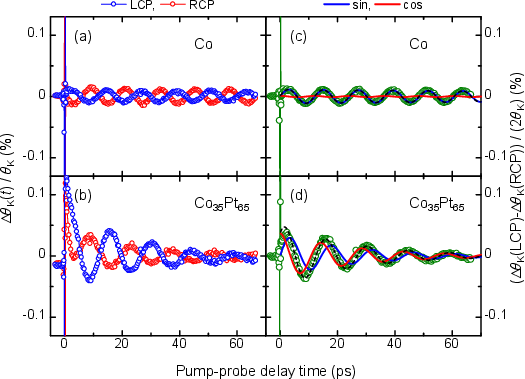}
    \end{center}
    \caption{(a) and (b) Circularly-polarized laser pulse induced change in normalized Kerr rotation angle ($\Delta \theta _{\rm K}(t) / \theta _{\rm K}$) in Co and Co$_{35}$Pt$_{65}$ alloy films with different photon-helicity using LCP and RCP laser pulses. 
    (c) and (d) Helicity-dependent change in normalized Kerr angle rotation ($(\Delta \theta _{\rm K}({\rm LCP})-\Delta \theta _{\rm K}({\rm RCP}))/(2\theta _{\rm K})$) as a function of time for Co and Co$_{35}$Pt$_{65}$ alloy films extracted from (a) and (b). 
    Broken and solid curves are results fitted using a sinusoidal decayed function [Eq. (2)].}
    \label{f2}
\end{figure}

Five-nanometer-thick Co$_{1-x}$Pt$_{x}$ alloy thin film samples with different compositions ($x$) were prepared by magnetron sputtering. 
The stacking structure of the sample is as follows: Si/SiO$_{2}$ sub./ Co$_{1-x}$Pt$_{x}$ (5) / Al (3) (thickness is in nanometers). 
The wavelength and pulse duration \SI{of the pump and probe laser pulse} used for the circularly-polarized laser induced magnetization dynamics measurements were 800 nm and 100 fs, respectively. 
The pump fluence ($F_{\rm p}$) used for the measurement was fixed at 4.6 J/m$^{2}$. 
An in-plane external magnetic field of 2 T was applied during the measurement.
\SI{The measurement was performed at room temperature.}
Figure 2 shows the effect of left circularly-polarized (LCP) and right circularly-polarized (RCP) laser on normalized Kerr rotation angle $\Delta \theta_{\rm K}⁄\theta_{\rm K}$ to determine the impact on magnetization dynamics in Co [Fig. 2(a)] and Co$_{35}$Pt$_{65}$ alloy [Fig. 2(b)].
Magnetization precession was excited after irradiation with a femtosecond laser pulse. 
The oscillation phase reversed when the photon-helicity was reversed, indicating a photon-helicity induced torque on magnetization. The slightly different amplitudes between the irradiation of the LCP and RCP resulted from thermally induced magnetization precession, likely caused by a slight misalignment between the sample and the applied field.
To extract the photon-helicity induced signal, the normalized difference signal ($(\Delta \theta_{\rm K} ({\rm RCP})-\theta_{\rm K} ({\rm LCP}))⁄(2\theta _{\rm K}) $) was taken as shown in Figs. 2(c) and 2(d). 
This photon-helicity-induced normalized change in the Kerr rotation angle was then fitted by the following equation:
\begin{align}
    &f(t)= \notag \\
    &\hspace{0.2cm}\left( \delta m_{\rm z, sin}\sin (2\pi f_0 t) + \delta m_{\rm z, cos}\cos(2\pi f_0 t)\right) \exp (-t/\tau _0) , \label{eq:fit}
\end{align}
where $\delta m_{\rm z, sin(cos)}$, $f_0$, and $\tau _0$ are sine (cosine) amplitude, frequency, and life-time of magnetization precession, respectively.
The broken and solid curves in Figs. 2(c) and 2(d) represent the fitted results based on Eq. (\ref{eq:fit}). 
In the case of Co, the magnetization precession excited by the photon-helicity was sine-like, indicating that the field-like torque was dominant. 
In contrast, a phase shift and cosine-like components, which are signatures of the damping-like torque, were exhibited by the Co$_{35}$Pt$_{65}$ alloy.

\begin{figure}[t!]
    \begin{center}
    \includegraphics[width=8.5cm,keepaspectratio,clip]{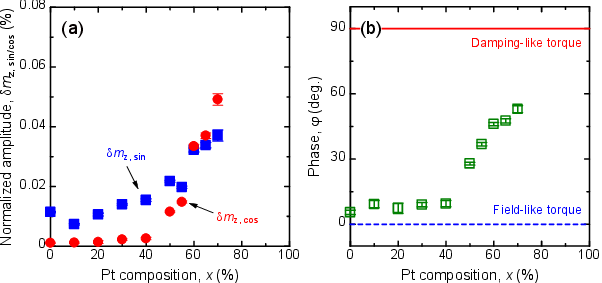}
    \end{center}
    \caption{ (a) Photon-helicity induced magnetization precession amplitude ($\delta m_{\rm sin}$ and $\delta m_{\rm cos}$) as a function of the Pt composition ($x$) in Co$_{1-x}$Pt$_{x}$ alloy. 
    Square and circle symbols are sine and cosine amplitude, respectively. 
    (b) Phase of magnetization precession ($\varphi $) as a function of $x$ extracted from (a). }
    \label{f3}
\end{figure}

Figure 3(a) shows $\delta m_{\rm z, sin}$ and $\delta m_{\rm z, cos}$ plotted as functions of the Pt composition ($x$). 
Both values increased with increasing $x$. In particular, a relatively significant increase of $\delta m_{\rm z, cos}$ was observed with an increase in $x$, compared with that of $\delta m_{\rm z, sin}$. 
The relative magnitudes of two values significantly changed, resulting in a change in the initial phase $\varphi $ (= $\tan ^{-1} (\delta m_{\rm z, cos}/\delta m_{\rm z, sin})$) [Fig. 3(b)].
This phase change indicates that the damping-like torque enhances with increasing Pt concentration in Co$_{1-x}$Pt$_{x}$, and therefore the SOC correlates with the above-mentioned physical scenario relevant to the OAM.

For a more quantitative discussion of both amplitudes ($\delta m _{\rm z}$) and phase change ($\varphi $), we consider the physics based on the simple dynamical model of magnetization (${\bf m}$) and the induced OAM ($\delta {\bf L}$). 
\SI{Although ab-initio theory of the inverse Faraday effect has been reported\cite{Berritta2016}, the theory was calculated in a few compound and derived to calculate OAM parallel to light propagation as well as magnetization direction for ferromagnets, which is different from the current experimental configuration. 
Therefore, we develop the model which allows us to consider precession of $\delta {\bf L}$ around magnetization with considering $\delta {\bf L}$ generation owing to absorption of angular momentum where the framework of OAM generation was discussed in\cite{Scheid2023}.}
\SI{The torque equation for $\delta {\bf L}$ was derived from the Heisenberg equation of motion (See Sec. I-A in Supplemental Material\cite{Sup}) in addition to OAM generation term and relaxation term as}:
\SI{
\begin{align}
    \frac{{\rm d}\delta {\bf L}}{{\rm d}t}={\bf Q}_{\rm L}-\xi ^{\prime }\delta {\bf L}\times {\bf m} -\frac{\delta {\bf L}}{\tau }, \label{eq:dynL}
\end{align}}
where ${\bf Q}_{\rm L}$ is the generation rate of the electronic OAM and its sign depends on the photon-helicity [Eq. (\ref{eq:Mp})].
Note that the form of ${\bf Q}_{\rm L}$ in Eq. (\ref{eq:dynL}) is based on absorption effect, {\it i.e.}, if there is no relaxation term OAM is preserved even after laser pulse passes away from the matter.
As discussed in Fig. 1, we assume the relaxation time ($\tau $) for the non-equilibrium OAM created and introduce SOC strength $\xi ^{\prime }$ as the frequency dimension for the coupling between magnetization ${\bf m}$ and $\delta {\bf L}$. 
\SI{The time-scale of dynamics induced by SOC and the relaxation process are shorter than pulse duration, thus we consider quasi-stationaly state during the pulse duration. 
The steady state condition ${\rm d}(\delta {\bf L})⁄{\rm d}t$ = 0 leads to a following equation}:
\begin{align}
    \delta {\bf L} &= \frac{\tau }{(\xi ^{\prime }\tau )^2 +1}{\bf Q}_{\rm L} - \frac{\xi ^{\prime }\tau ^2 }{(\xi ^{\prime }\tau )^2 +1}{\bf Q}_{\rm L} \times {\bf m} \notag \\
    &\hspace{1cm}+ \frac{(\xi ^{\prime }\tau )^2 \tau }{(\xi ^{\prime }\tau )^2 +1}({\bf Q_{\rm L}}\cdot {\bf m}){\bf m}. \label{eq:dL}
\end{align}
\SI{The first term represents the quasi-static electronic OAM whose direction is parallel to the light propagation direction and the second term arises from the rotation of OAM around ${\bf m}$ due to SOC between OAM and magnetization.}
The low-frequency dynamics of magnetization in ferromagnetic metals can be expressed using the Landau-Lifshitz equation with torque ${\bf T}$ as follows:
\begin{align}
    \frac{{\rm d}{\bf m}}{{\rm d}t} = -\gamma \mu _0 {\bf m}\times {\bf H}_{\rm eff} + {\bf T}, \label{eq:LLG}
\end{align}
where, $\gamma $, $\mu _0$, and ${\bf H}_{\rm eff}$ are the gyromagnetic ratio, vacuum permeability, and effective magnetic field, respectively.
Here, we assume the torque is due to the SOC \SI{${\bf T}=\frac{\gamma \xi ^{\prime }}{\mu _{\rm at}} {\bf m}\times \delta {\bf L}$ (See Sec. I-A in \cite{Sup})}, where $\mu _{\rm at}$ is the atomic magnetic moment. 
Then, we derived the following dynamical equation:
\begin{align}
    \frac{{\rm d}{\bf m}}{{\rm d}t} &= -\gamma \mu _0 {\bf m}\times {\bf H}_{\rm eff} +\frac{\gamma }{\mu _{\rm at}}\frac{\xi ^{\prime }\tau }{(\xi ^{\prime }\tau )^2+1}{\bf m}\times {\bf Q}_{\rm L} \notag \\
    &\hspace{0.2cm} -\frac{\gamma }{\mu _{\rm at}}\frac{(\xi ^{\prime }\tau )^2}{(\xi ^{\prime }\tau )^2+1}{\bf m}\times ({\bf Q}_{\rm L}\times {\bf m}). \label{eq:torque}
\end{align}
The second and third terms are the field- and damping-like torques, respectively, and both are governed by the time scale of $\xi ^{\prime }$ and $\tau $, as discussed in Fig. 1. 
Here, laser was irradiated from the film normal; thus, ${\bf Q}_{\rm L}$ = -$Q_{\rm L} {\bf z}$. 
Then, the field- and damping-like torques were along ${\bf y}$ and ${\bf z}$ directions. 
The \SI{z-component} magnetization precession amplitudes can be derived by \SI{impulsive response due to the torque ${\bf T}$} as\SI{\cite{Iihama2022}},
\begin{align}
    &\delta m_{\rm z, sin} = \frac{\gamma }{\mu _{\rm at}}\frac{\xi ^{\prime }\tau }{(\xi ^{\prime }\tau )^2 + 1}\beta Q_{\rm L}\Delta t , \label{eq:dmsin} \\
    &\delta m_{\rm z, cos} = \frac{\gamma }{\mu _{\rm at}}\frac{(\xi ^{\prime }\tau )^2}{(\xi ^{\prime }\tau )^2 + 1} Q_{\rm L}\Delta t , \label{eq:dmcos}
\end{align}
where $\Delta t $ is the laser pulse duration.
Here, we use ${\bf H}_{\rm eff}=H {\bf x} - M_{\rm eff}({\bf m}\cdot {\bf z}){\bf z}$ with an external magnetic field ($H$) and effective demagnetizing field ($M_{\rm eff}$). 
The factor $\beta =\sqrt{H/(H+M_{\rm eff})}$ is related to the elliptical shape of magnetization precession due to the demagnetizing field in the thin film.

\begin{figure}[ht]
    \begin{center}
    \includegraphics[width=8cm,keepaspectratio,clip]{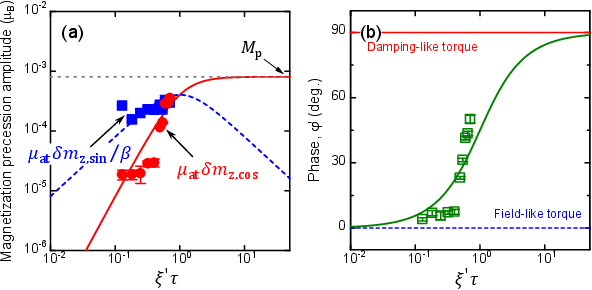}
    \end{center}
    \caption{ 
        (a) Magnetization precession amplitude ($\mu _{\rm at}\delta m_{\rm z, sin}/\beta $ and $\mu _{\rm at}\delta m_{\rm z, cos}$) and (b) \SI{phase ($\varphi ^{\prime }_{\rm exp}= \tan ^{-1} (\beta \delta m_{\rm z, cos}/\delta m_{\rm z, sin})$)} plotted as a function of $\xi ^{\prime }\tau $ in the model of photon-helicity induced OAM transfer torque. Broken and solid curve in (a) are results calculated based on $M_{\rm p} \xi ^{\prime }\tau ⁄(1+(\xi ^{\prime }\tau )^2 )$ and $M_{\rm p} (\xi ^{\prime }\tau )^2 ⁄(1+(\xi ^{\prime }\tau )^2 )$, respectively, where $M_{\rm p}$ corresponds to the orbital magnetization generated by light.
        Solid curve in (b) is the result calculated based on \SI{$\varphi ^{\prime }_{\rm the} = \tan ^{-1 }(\xi ^{\prime }\tau )$}.
        Here, $\xi $ scales with $Z_{\rm ave}^2$ and orbital relaxation time $\tau $ of 1 fs.
    }
    \label{f4}
\end{figure}

The experimental and theoretical amplitudes are compared in Fig. 4(a). 
In this figure, we plot the normalized values of experimentally-obtained magnetization precession amplitudes ($\mu _{\rm at} \delta m_{\rm z,sin}⁄\beta $ and $\mu _{\rm at} \delta m_{\rm z,cos}$) as functions of $\xi ^{\prime }\tau $.
Here, we evaluated $\xi ^{\prime }$ for Co-Pt alloy with different compositions ($x$) using the values of $\xi ^{\prime }$ interpolation of the SOC strength energy for Co and Pt (Co: 0.085 eV, Pt: 0.72 eV)\cite{Sipr2008} with the well-known scaling rule $\sim Z_{\rm avg}^2$\cite{Shanavas2014}, where $Z_{\rm avg}$ is the average atomic number of Co-Pt alloy.
We also set approximately used fixed value of $\tau $ $\sim $ 1 fs, which approximates photo-exited electron relaxation time, {\it i.e.}, $\tau $ $\sim $ 1 fs for Co\cite{Bauer2015}. 
\SI{(See Sec. I\hspace{-1.2pt}V in Supplemental Material for more detailed discussion for $\xi ^{\prime }$ and $\tau $ for the alloys\cite{Sup})}.
The theoretical curves in Eqs. (\ref{eq:dmsin}) and (\ref{eq:dmcos}) are shown with the broken and solid curves, respectively, in Fig. 4(a) under the assumption of a fixed value of $M_{\rm p}$ = $\gamma Q_{\rm L} \Delta t$ $\sim $ 0.8 $\times $ 10$^{-3}$ [$\mu _{\rm B}$]. 
The theoretical magnetization precession phase is given by the ratio between the field- and damping-like torques in Eq. (6):\SI{
\begin{align}
\varphi _{\rm the}^{\prime } = \tan ^{-1} (\xi ^{\prime }\tau ). \label{eq:phase}
\end{align}
}
\SI{Figure 4(b) shows the theoretical and experimental phases of magnetization precession as a function of $\xi ^{\prime }\tau $. 
Here, experimental phase is corrected by the factor $\beta $, {\it i.e.}, $\varphi _{\rm exp}^{\prime } = \tan^{-1}(\beta \delta m_{\rm z, cos}/\delta m_{\rm z, sin}) $ from Eqs. (\ref{eq:dmsin}) and (\ref{eq:dmcos}) to compare with the theoretical phase $\varphi _{\rm the}^{\prime }$ for the torques [Eq. (\ref{eq:phase})]}
In those figures, the experimental data for the amplitude and phase are in good agreement with those obtained from the theoretical model with the same footing.

The pertinent physics discussed thus far for helicity-induced magnetization dynamics is worth noting. 
As mentioned in the introduction, the inverse Faraday effect is traditionally considered to effectively generate magnetic fields in both magnetic metals and dielectrics \SI{owing to the magneto-optical coupling\cite{Kirilyuk2010}}. 
Thermodynamic considerations indicate that the effective magnetic field generated by the \SI{magneto-optical coupling} is proportional to the Faraday or Kerr effect, which is characterized by magneto-optic constant $Q$. 
However, our sample showed {\it a reduction of the Kerr effect with an increase in the Pt concentration}, contrary to the significant increase in helicity-induced torque with increasing Pt concentration\SI{(See Sec. I\hspace{-1.2pt}I-A and V-A in Supplemental Material\cite{Sup})}.
\SI{Thus, significant increase of both field- and damping-like torques with Pt concentration cannot be explained by the inverse Faraday effect due to magneto-optical coupling.}
\SI{On the other hand,} recent microscopic theories facilitated further understanding of the nature of field- and damping-like torques in metals via SAM directed to $\delta {\bf S}$ and $\delta {\bf S}\times {\bf m}$, respectively, where $\delta {\bf S}$ is SAM induced by the circularly-polarized light\cite{Freimuth2016}. 
The theoretical computations predicted that field-like torque was significantly enhanced in ferromagnetic metals with large SOC such as FePt, with the damping-like torque being much smaller than the field-like torque. 
\SI{Concerning the SAM generation}, the emergence of the damping-like torque with increasing Pt concentration is difficult to be interpreted as the SAM generation in ferromagnets, {\it i.e.}, ${\bf M}_{\rm p}\sim \delta {\bf S}$ in Eq. (\ref{eq:Mp}), similar to that observed in GaMnAs\cite{Nemec2012}.
When $\delta {\bf S}$ is generated by circularly polarized light irradiation, the time-scale is characterized by the exchange coupling ($J_{\rm ex}$) of the induced $\delta {\bf S}$ and ${\bf m}$, and relaxation time of the induced spin ($\tau _{\rm S}$). 
Hence, we can expect $\varphi \sim \tan^{-1}(J_{\rm ex} \tau _{\rm S}⁄\hbar )$, similar to Eq. (\ref{eq:phase}).
It is difficult to consider that the phase increases with increasing Pt concentration ($x$) in Co-Pt alloys because Pt addition would deteriorate ferromagnetism, that is reduction of $J_{\rm ex}$.
\SI{Also, Pt addition reduces $\tau _{\rm S}$ due to its strong SOC according to the Elliot-Yafet mechanism\cite{Zutic2004}}. 
Similarly, the damping-like torque has been discussed in terms of the SAM induced in the heavy-metals in ferromagnet / heavy-metal bilayers; however, the physics behind this torque is related to the spin-transfer-torque owing to the SAM flow in the bilayer\cite{Choi2020}. 
\SI{Moreover, ultrafast damping-like torque was observed in antiferromagnet\cite{Tzschaschel2020}.
This was caused by exchange-enhancement of the Gilbert damping torque which may be a unique feature in some antiferromagnets and not the case in ferromagnets studied here (See Sec. V-B in Supplemental Material\cite{Sup} for detailed discussion).
}
Thus, we ruled out the physical mechanisms of the damping-like torque proposed to date.

Our study indicates that the OAM induced by light plays a much more important role than the SAM induced by the light. 
We \SI{try to quantify} generation of \SI{OAM} and orbital magnetization owing to light absorption in metals\cite{Scheid2023}. 
\SI{The calculation of $M_{\rm p}$ generated due to conservation of angular momentum leads to a value $\sim $ 10$^{-2}$ [$\mu _{\rm B}$] one order of magnitude larger than that used to explain the experimental result $\sim $ 10$^{-3}$ [$\mu _{\rm B}$] (Fig. 4(a) and see Sec. V\hspace{-1.2pt}I in Supplemental Material\cite{Sup} for detailed calculation)}. 
\SI{This fact indicates that some part of the induced OAM contributes to the magnetization torque.
It is likely that the torque may dominantly be caused by the induced OAM with $d$-electron character since magnetization is mostly composed of the $d$-electron.
On the other hand, conservation law accounts induced OAM for $sp$-electrons as well as that for the $d$-electron, which might be a source of the difference between experiment and calculation based on the conservation law.
Thus, for more quantitative discussion, we need the quantum theory of the angular momentum dynamics and torque induced by light taking account of realistic band structure beyond the naive phenomenological model discussed here, which is out-of-scope of the Letter and will be subjects for future works}.

In summary, we investigated helicity-dependent laser-induced magnetization precession in Co$_{1-x}$Pt$_{x}$ alloys. 
We demonstrated that the magnitude and the phase of the magnetization precession induced by the circularly-polarized laser pulse systematically increase with the Pt concentrations in the Co$_{1-x}$Pt$_{x}$ alloy film. 
Those evolutions are attributed to two different types of induced torques, field- and damping-like torques. 
The experimental data are well described by a simple model that considers the electronic OAM generated by light and its coupling with magnetization via SOC. 
This study deepens the understanding of the physics relevant to the interaction with photon-helicity and matter. 
Furthermore, this work opens alternative route for manipulating magnetization in magnetic substances.

This work was supported by Grant-in-Aid for Scientific Research (Nos. 21H04648 and 21H05000), Grant-in-Aid for Challenging Research (No. 24K21234), Grant-in-Aid for Transformative Research Areas (No. 24H02235), JST PRESTO (No. JPMJPR22B2), X-NICS of MEXT (No. JPJ011438), CSIS cooperative research project in Tohoku University, the Asahi Glass Foundation, the Murata Science Foundation, Advanced Technology Institute Research Grants, FRIS Creative Interdisciplinary Collaboration Program in Tohoku University. This work was also supported by the ANR-20-CE09-0013 UFO, by “Lorraine Université d'Excellence” reference ANR-15-IDEX-04-LUE, and by the French National Research Agency through the France 2030 government grants PEPR Electronic EMCOM (ANR-22-PEEL-0009).  S. I. thanks to TI-FRIS fellowship at Tohoku University. K. N. and K. I. thank to GP-Spin at Tohoku University. S. M. thanks to CSRN of CSIS at Tohoku University. The authors thank Gyung-Min Choi for valuable discussion.

\bibliography{export2}

\end{document}